# A formative measurement validation methodology for survey questionnaires

Mark Dominique Dalipe Muñoz
*Mathematics Department*
*(Iloilo Science and Technology University – Main Campus)*
Iloilo City, Philippines
markdominique.munoz@students.isatu.edu.ph

*Abstract:* Model misspecification of formative indicators remains a widely documented issue across academic literature, yet scholars lack a clear consensus on pragmatic, prescriptive approaches to manage this gap. This ambiguity forces researchers to rely on psychometric frameworks primarily intended for reflective models, and thus risks misleading findings. This article introduces a Multi-Step Validation Methodology Framework specifically designed for formative constructs in survey-based research. The proposed framework is grounded in an exhaustive literature review and integrates essential pilot diagnostics through descriptive statistics and multicollinearity checks. The methodology provides researchers with the necessary theoretical and structural clarity to finally justify and adhere to appropriate validation techniques that accurately account for the causal nature of the constructs while ensuring high psychometric and statistical integrity.

## 1. INTRODUCTION

Pilot testing procedures are a necessary foundational process to ensure the quality and integrity of survey questionnaires [15,23,25], which serve as the primary tools for gathering observational data across fields such as Marketing, Information Systems, and Management [1,3]. However, from my prior experience on the current academic practices, I have frequently encountered persistent methodological misconceptions and incorrect practices that require immediate academic intervention to achieve greater rigor and confidence in the corresponding findings.

This methodological gap is not new and has been well-documented by previous researchers throughout history when the psychometric methods were gaining scientific adoption in the 19[th] century across various social disciplines [1,5]. For instance, researchers in management and social science fields often "force" their constructs to fit within the reflective model framework regardless of their actual theoretical nature [1,4], which functionally misleads their true psychometric nature. Evidence suggests that their historical roots between the reflective and formative measurement models are key reasons for the disparity. Formative constructs have been overshadowed by the continuous development of psychometric tools intended for reflective constructs [1,5,6] as one measure from this said framework is the widespread use of Cronbach's alpha for assessing instrument reliability due to its computational simplicity and adoption across statistical packages [17].

However, relying solely on Cronbach's alpha as a measure for assessing instrument quality overlooks a fundamental distinction in measurement theory. While current literature already acknowledges the limitations of using Cronbach's alpha alone [14,16,17,18,19] and have thus recommended to explore other types of reliability such as test-retest, split-half, and interrater and types of validity such as face, construct, and content [5], the subtle gap is on the very nature of the constructs we seek to measure and our pre-specified items or questions. We are missing a critical point that Cronbach's alpha, as well as these reliability and validity types, are rooted deeply in Classical Test Theory (CTT) [3,5]. They are necessary diagnostic tools only when the construct is assumed to be reflective, where each individual item is viewed to conceptually represent the construct in a consistent manner [3]. This is analogous to assuming the items are multiple ways to represent a construct and then finding a common ground to cancel out the item-based variants, and making the composite measure more robust. These items are expected to be highly correlated because they are assumed to share a common cause, satisfying stringent assumptions such as tau ($\tau$)-equivalence and unidimensionality [16,18].

The psychometric properties between reflective and formative constructs are seen as opposites of one another [2]. The indicators on formative models are viewed as "forming together" into a unified construct of interest, and any changes to the combinations of the indicators fundamentally alter the meaning of the construct [1,3,7]. Their relationship is best expressed as a linear composite, often modeled using regression techniques like Partial Least Squares Structural Equation Modeling (PLS-SEM) [2]-[3]. Because formative indicators do not share a common cause, internal consistency reliability is not required and can, in fact, be problematic as it can lead to multicollinearity or redundancy of items and unstable indicator weights [1].

Misspecified constructs pose a risk of obtaining misleading and biased statistical results [1]-[2]. Researchers would sometimes grapple with the meaning of low Cronbach's alpha values even when the subject matter experts (SMEs) have validated the relevance of each item to the construct's domain. It does signal a psychometric problem of inconsistency when the constructs are reflective, but it can threaten the content validity of a formative construct since ensuring that the indicators adequately represent the scope of the construct is its key characteristic [3].

While the distinction between reflective and formative constructs and their corresponding validity and reliability protocols is well-documented in the context of Structural Equation Modeling (SEM) [1,3,13,22] when assessing the outer (measurement) model, there is still a need to emphasize the key distinctions of these two types of constructs outside of SEM, especially for the purposes of questionnaire validation. This article thus aims to address a long-standing gap in scale development by proposing a formal validation protocol among researchers during pilot testing phases.

In this article, the operational definition of three main terms: construct, indicator, item, and model are addressed to provide clearer comprehension. A *model* will refer to a framework or blueprint and thus, a formative model meant the overall system that the construct, indicator, and other parameters interrelate with one another while assuming the formative context. Now indicator and item are used interchangeably but for *indicator* is meant as just a more niche or formal term similar to what a questionnaire *item* is. Finally, a *construct* is something that is represented by the items either as a representation or a contributor. When referring to a reflective construct, it always implies the indicator or the item is reflective and vice versa.

## 2. FORMATIVE MEASUREMENT MODEL

The formative model is a psychometric model that explains the nature of the relationship between the construct and indicators by assuming that the indicators collectively form the meaning of the underlying construct [2,4,29], yet it is often less understood and acknowledged than the reflective model. Being associated with index construction, its origins are traced back to the principles of operationalism, which posits that a concept is directly represented or measured from the measure itself [4]. Its entire meaning depends on the nature of the indicators themselves. Hence, $\eta$ represents the latent variable and x is an indicator, then

$$\eta \equiv x \qquad (1)$$

This concept can later be generalized into multiple indicators. Suppose we have indicators $x_i$ where $i = 1,2,…,n$, then these $n$ indicators collectively form the inherent meaning of the latent variable $\eta$. Its meaning thus changes depending on the nature of $x_i$ involved. A formative specification can be implied as [4,7]:

$$\eta \equiv \gamma_1 x_1 + \gamma_2 x_2 + … + \gamma_n x_n \qquad (2)$$

where $\gamma_i$ is a parameter reflecting the unique contribution of $x_i$ in explaining $\eta$. These terms are analogous to what we commonly known as the weights of each indicator.

Bollen and Lennox [4] also provided another specification as follows:

$$\eta \equiv \gamma_1 x_1 + \gamma_2 x_2 + … + \gamma_n x_n + \zeta \qquad (3)$$

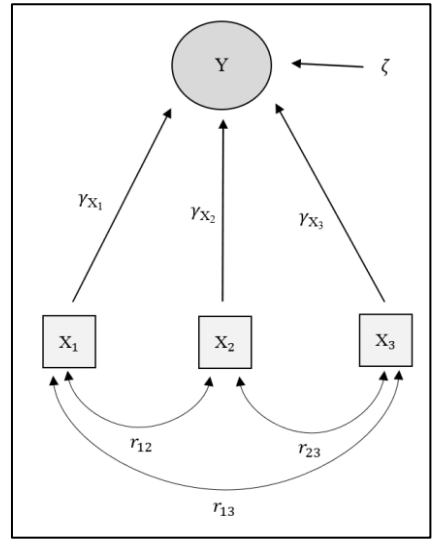

**Fig. 1.** *Formative measurement model illustration (Adapted from [4])*

Equation 1 captures the widespread use of single-item measures during the 1960s-70s, yet it limits the possibility of multiple measures to represent the same construct. For equations 2 and 3 [4], we can see that the only difference with Bollen and Lennox's specification [4] is the introduction of a disturbance term $\zeta$. This term defines the remaining aspect that has been left unrepresented by any of the indicators, which would ideally make the collective representation of the indicators theoretically "complete". Equation 3 emphasizes that the construct is psychometrically represented as a linear composite of the weighted contributions of each indicator. It is often an imperfect representation because of the abstract nature of the construct that lacks a defined scope. From a psychometric perspective, Diamantopoulos and Wiklhofer [4] presented four key characteristics that distinguish its nature from reflective constructs.

Property 1.
$$\boldsymbol{\eta = f(x_i)}$$

The inherent meaning of the formative construct is heavily dependent on the nature of the indicators as a whole and across each individual indicator.

Property 2.
$$\boldsymbol{Cov(x_j, x_k) = \sigma_{jk} \text{ where } j, k \in \{1, 2, \dots, n\}}$$

The indicators are assumed to cause and therefore precede a latent construct, correlations (or covariances) among indicators guide the selection of indicators that are to be retained or removed. Since each correlation value is unique, which directly informs the relative redundancy between any two arbitrary indicators alone, we don't expect any particular trend to surface in these correlation values since the implications of the model do not depend on them. Hence, we can identify $\sigma_{jk}$ here as a free parameter that can take on any values (positive, negative, or zero).

Property 3.
$$\boldsymbol{Cov(x_i, \zeta) = 0}$$

Indicators do not have "error" terms associated with the indicators because each indicator is assumed to be a precise measurement of a construct (a direct consequence of Equation 1). Instead, error variance is represented only in the disturbance term $\zeta$ associated with the latent construct.

Property 4. $\boldsymbol{df = Number\ of\ unique\ data\ points - Number\ of\ free\ parameters}$

This condition is characteristic when a formative model is part of the structural equation modeling (SEM) framework. A single formative construct on its own is statistically underidentified. In the context of SEM, model underidentification means that the number of unique data points is less than the number of free parameters, resulting in insufficient information to identify a solution to a set of structural equations [2]-[3]. Simply put, multiple possible solutions are found but unstable for a given environment, and thus, the model might prefer to have multiple constructs to compensate for the insufficient unique data points, yet this is not always the case. The nature of formative specification poses a problem because the error variances are associated with the constructs that are fewer in quantity than the indicators [3]. Our current statistical approaches, however, circumvent or avoid this modeling problem because we are not estimating any "solution" on the formative model in the form of empirical weights. Rather, we will use theoretical weights for each indicator which is tied to literature review and expert feedback.

From these conditions, we can see that internal consistency reliability and construct validity in terms of convergent and discriminant validity are not suitable even in its fundamental model-based characteristics because the constructs are viewed not as scales but as indices formed from a composition of its indicators.

## 3. GUIDANCE ON REFLECTIVE AND FORMATIVE MODELS

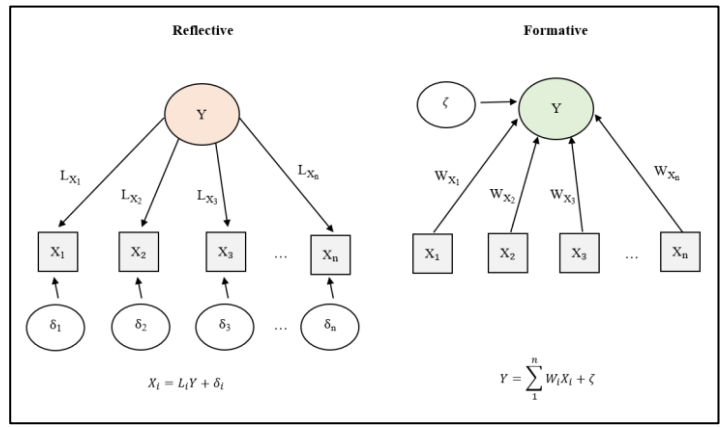

**Fig. 2.** *Comparative illustration between reflective and formative models*

We can now refer to a general guide of the distinction between reflective and formative constructs to aid the researchers in pre-specifying each construct and therefore, have better guidance on the right psychometric approaches to use. Reflective and formative models are also known as principal factor models and latent variable models, respectively.

| Characteristic | Reflective Model | Formative Model |
| --- | --- | --- |
| Causality of construct | Construct → Indicator (Construct causes the indicator) | Indicator → Construct (Indicator causes the construct) |
| Nature of Error | Measurement Error Term (Measures the discrepancy on the ability of an item to represent the construct) | Disturbance Term (Measures the unrepresented aspect of the construct relative to its the overall theoretical scope) |
| Conceptual relationship among items | All items are conceptually related with one another. | No strict requirement of conceptual relatedness across any two items. |
| Domain of items | Items are simply a sample of representations of a construct. | Items encompass the entire scope of representation of a construct. |
| Content validity | A useful validity check to ensure relevance of the items. | A mandatory validity check to ensure the integrity of the construct. |
| Covariance among items | Collinearity among items is expected. | No expectation of collinearity. High collinearity poses problems. |
| Internal consistency | Required. | Not required and problematic because of redundancy. |
| Construct validity | Required especially on convergent and discriminant validity. | Only external construct validity (Its relationship towards other constructs) if appropriate. |

| Mismatch on Theoretical and Empirical Findings | Rare and safe case because items are simply manifestations of an already established construct. | Problematic and likely when the existing literature does not match the expectations from the statistical findings. This is evident when estimating empirical weights after data analysis. |

**Table 1.** *Comparison between reflective and formative models (Adapted and synthesized from [1] – [2])*

Jarvis et al. [1] also proposed the general guidelines on specifying whether a construct is formative or reflective of which the selection should rely on the discretion of the researchers and/or Subject Matter Experts (SMEs). Since the original table regarding this concept appears to be conceptually overlapping from our previous table in this study, a prescriptive version of its content is provided as follows:

Ideally, there are three aspects you have to assess on formative scale development [2]: (1) Construct: Direction of causality, (2) Indicators: Interchangeability and covariation, (3) Extraneous factors on indicators: Nomological net of the construct indicators.

*a. Direction of causality*

Identifying the causal direction of each construct should rely on how it is defined on existing literature about the construct itself and its related phenomena [4,30]. Prioritize key articles that are directly utilized to the theories you establish at the beginning of your study as research frameworks, rather than simply an arbitrary pool of articles [29,30]. This avoids the perception bias from "niche" studies because researchers form their own understanding about a particular topic or theory that can influence how they define the term in the context of their study.

It is also important to acknowledge that the definition of terms is often intended for the particular context of a study given the quantitative nature of scale development, which can limit its direct application across other studies. Pay attention on how a construct is defined from your literature sources. Does the construct imply an underlying abstract concept that you can measure by itself? If so, it is reflective. Otherwise, it is formative if the definition of the construct directly depends on a combination of multiple factors or aspects which are explicitly stated. However, the researchers might come across a construct that can be framed as both reflective or formative depending on the corresponding literatures that support it. In this case, they are given the freedom to choose the model that is more relevant to the objectives of the study.

*b. Interchangeability and covariation among indicators/items*

After pre-specifying the anticipated psychometric model of each construct, the researcher will create or generate the appropriate survey items. If the researchers happened to have access to a standardized questionnaire from a previous study, it is recommended to immediately assign all constructs as reflective. For formative constructs, the gold standard is for the researchers to deliberately generate an item pool [10] based on the existing literature, expert feedback from Subject Matter Experts (SMEs), and assumed psychometric model for the underlying construct. This is more commonly known as "self-made questionnaire" because the construct's meaning heavily depends on the local context of the researchers and the respondents.

The key hint to look first is for whether the absence of each item would alter the substance of the construct. This is the interchangeability characteristic of items. If it does alter, the items are formative since each item has a unique contribution towards the construct, making them non-interchangeable. If it doesn't, it's reflective.

However, if the researchers thought the distinction is insufficient or ambiguous from current examination of their constructs, we assess the second aspect: covariation. This refers to the tendency of an item to covary with the trend of another item, and is a key characteristic on reflective constructs, but can also be found on formative items. Hence, the distinction is on the necessity to covary. If responses across items should be expected to follow a similar trend with one another, then it is not formative. This characteristic is analogous to internal consistency reliability where you expect the responses to consistently align with the rest of item responses on their corresponding construct. Because if you observed that a respondent can theoretically disagree on one item but agree on another, it's a classic suspicion of a formative indicator.

What if we found that the items we've generated are actually a mix of formative and reflective constructs? We must align to what your construct actually assumes at the first place (i.e. If the construct is formatively defined, you must look for formative indicators). This ensures that we adhere to the very core psychometric characteristics that distinguish these two models since the nature of construct and indicator, by definition, are interconnected.

*c. Nomological net of construct indicators*

This refers to the external factors that either influence (cause) or be influenced (effect) by the indicators and is a direct extension from the covariation aspect [30]. If a factor causes or is affected one indicator, then it should also cause or be affected by another indicator under the same construct. This trait is characteristic of reflective constructs, however, since it aims consistency and stability of representing the construct. The assessment of the Nomological Net is related to the structural model that occurs after the measurement model has been purified in the context of Structural Equation Modelling [1,3]. In other words, it means that assessing Nomological Net only happens after completing the questionnaire validation process and the researchers are set to analyze the data gathered from the actual respondents. Hence, this aspect will be excluded from our framework in this study.

## 4. HIGHER-ORDER MODELS

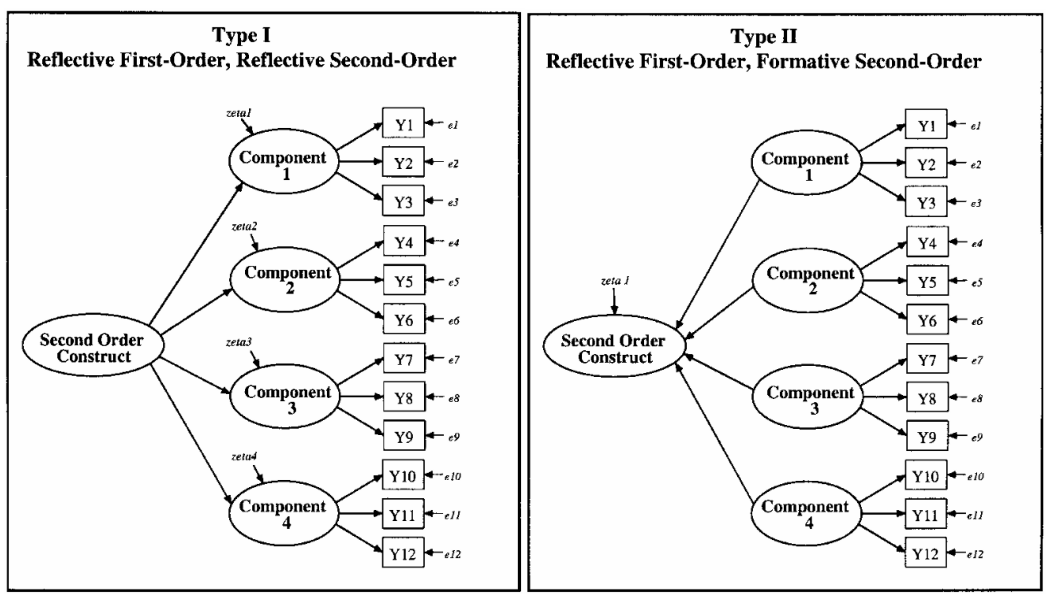

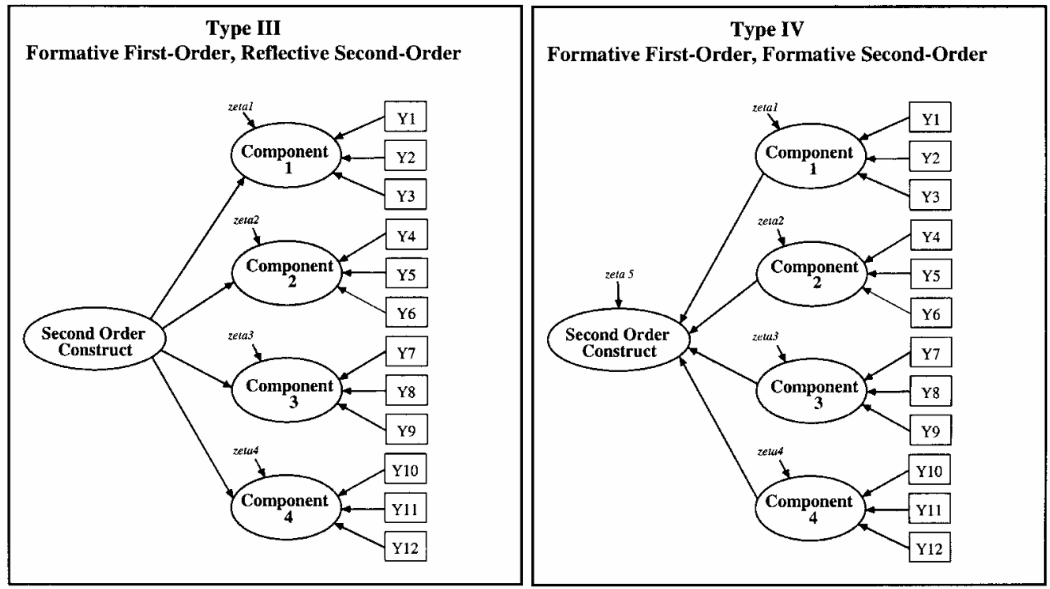

**Fig. 3.** *Variants of higher-order models [1].*

So far, we have only encountered the intricacies between the two models on a single order? But what about for higher level models? There are different combinations of these models involving a mix of both formative and reflective constructs [1]. This challenges the traditional notion of a construct that should be conceptually and empirically unidimensional to be meaningful. Each of these models are still guided by existing literature and expert guidance where the previous literature review guidelines still apply. For the sake of discussion, let's assume for now a second-order model where the first-order constructs become the indicators of the second-order constructs.

While several modeling techniques are used to estimate the values of latent variables, they require a large sample size and for a pilot testing involving a small number of respondents, these tools may struggle to provide accurate estimation [1,3]. Hence, for pragmatic reasons, extract a composite measure of each construct by computing the weighted mean or median for each respondent. The weighted means now serve as the proxy data of the first-order constructs and will be used for further validation on the second order constructs. We will still use the similar diagnostic measures which will be presented in the later sections. This diagnostic approach can be generalized further beyond second-order models.

## 5. NATURE OF PILOT TESTING

Diamantopoulos and his colleagues [4] have identified four key issues to assess in the interest of an effective scale development, which will serve as the primary guide on what to assess during the pilot testing phase.

*a. Content Specification*

This refers to the scope of the latent variable as seen in the previous sections. Since it is important for formative constructs to represent its scope with as much breadth as possible from its causal indicators, having a clear and well-defined construct is crucial for a clearer and more informed decision on deciding the appropriate psychometric model. Researchers are recommended to refer back to (a) Direction of causality that presents key highlights on this step as well as its significance in the pilot testing process.

*b. Indicator Specification*

This refers to how the researchers should assign and create the corresponding items of each construct. It requires careful consideration of the underlying aspects that the construct may directly or implicitly imply. Again, I recommend to refer back to the (b) Interchangeability and covariation among indicators/items for a full description on this aspect.

*c. Indicator Collinearity*

This refers to the strength of relationship among any pair of indicators. It touches the diagnostic aspect of validating whether the pilot data will also adhere with the assumed theoretical context. Excessive collinearity leads to poor discrimination on the unique effects of each indicator and therefore assess its significance on measuring the latent variable [3]. We conduct its corresponding tests after the pilot data is obtained and will be performed iteratively until we obtain results that satisfy our empirical conditions.

*d. External Validity*

This refers to the extent of how our current formative constructs relate to other constructs. Ideally, this can be determined by examining the covariance of each indicator to another external variable. Since the external variable acts as a global construct that provides a valid baseline for our current indicators, indicators that correlate with this variable are retained because it strengthens our assumption that our indicators mirror an ideal representation of our construct. It can be related to nomological validity as it concerns relationships with external factors. Hence, like nomological validity, this issue is beyond the intended internal scope and will thus be excluded from our framework.

**6. FRAMEWORK PROPER**

Literatures surrounding the pilot testing process among survey questionnaires are well-documented. While its paradigm sufficiently lays a practical step by step guide for student researchers in survey questionnaires, they often revolve by assuming the constructs to be reflective. Churchill's validation methodology [6] is an eight-step process incorporating the use reliability and validity measures as default steps to ensure instrument quality, and also advocated the use of multi-item measures as single items lack the essential scope a construct needs for a more stable finding. We now seek to contextualize and refine these protocols on formative constructs to advocate its use whenever deemed necessary. I replaced the original reliability and validity measures with descriptive statistics and multicollinearity measures while acknowledging the iterative aspect of validating questionnaires. The purification step on his protocol which promoted the use of Cronbach's alpha and factor analysis to inspect the psychometric characteristics of the constructs is inappropriate for formative models [1,17]. Hence, the researchers are recommended to perform the following sections in particular order.

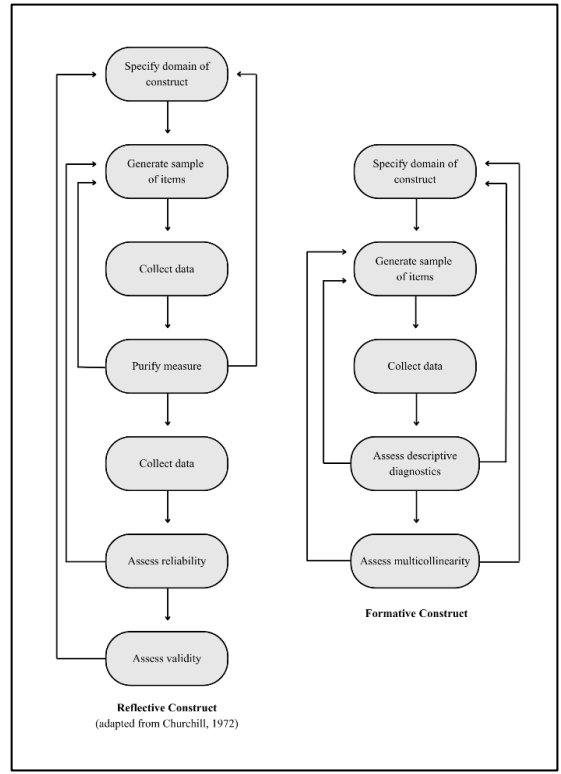

**Fig. 4.** *Validation flow between Reflective Construct (adapted from [6]) and Formative Construct.*

## 6.1 Domain of construct

The scale development process begins with identifying what and how the identified constructs will be measured based from previous literature and expert review [21,24,30]. This is where the researchers will decide and set each construct to assume either the formative or reflective psychometric model, but not both. Corresponding literature should then be cited to support the pre-defined psychometric models for each construct.

## 6.2 Content validity

Once the constructs are established, the researchers must create corresponding items under each construct from previous literature [24,30]. While there are common conventions of either choosing to deliberately formulate the items on their own (self-made) or utilize an existing standardized questionnaire from an existing study, I recommend the self-made option if one of your constructs is formative because of the theory-driven aspect of the items as it allows you to account the contextual nuance of your population and its relation to your indications. This is known as generating an item pool [8,10]. Adapting an existing questionnaire will risk not accounting for the aspects of your construct that are only unique in the context of your study [20].

The nature and number of indicators of each construct should be guided by literature and the related field of study [12,30]. If no clear guideline on the number of items exists, a heuristic rule of thumb of at least 5 items is recommended as a last resort, provided that the items will sufficiently represent the underlying aspects of the constructs found on existing literature. The researchers will then justify selection of items by citing the corresponding articles that either act as: (a) definitional source: define the construct and present the aspects, (b) mirror source: present the precise items used that match the needed aspects to measure. For (a), the researchers will highlight the aspects from the presented definition and reason a theoretical relevance of one or more aspects to an item's context. For (b), the researchers can simply use the item, provided that they have permission from the author of the study and have ensured the suitability of the item by assessing the population of the original study.

For the theoretical weighing, theoretical weights will be pre-defined as preliminary values for the pilot testing process in setting the relative importance of the items in measuring the construct. The weights seek to answer the question: "How important is each item in understanding the construct when compared with the rest of the items?" While the decision for setting the weights relies on either the researchers or subject matter experts (SMEs), I recommend SMEs where you can calculate Content Validity Ratio (CVR) values of each item $i$ based from their collective responses [8,9,24]. Not only would the CVR values could act as theoretical weights for the indicators, but the SMEs can also provide qualitative feedbacks on the questionnaire to correct or improve its scope. Having the researchers to pre-define the weights should be a last resort due to logistical constraints such as lack of availability among SMEs and limited time as it sacrifices methodological rigor. If it does, a score of an item can be calculated using the arithmetic average of the researchers' rating from 1- not essential to 5- essential which will rely on the corresponding literature that support its significance on measuring the construct.

$$CVR_i = \frac{n_e - \frac{N}{2}}{\frac{N}{2}} \qquad (4)$$

Where:

$n_e$: number of raters indicating "essential"

N: total number of raters

### 6.3. Update questionnaire and collect pilot data

This section includes steps 3 and 4 from the proposed framework. The refined questionnaire draft from section 4.2 can now be answered by the pilot respondents. Pilot testing will be conducted with two distinct objectives in mind:

(a) Pilot respondents will assess the face validity of the instrument by answering a supplementary questionnaire containing questions about the quality of instrument such as clarity and preciseness of language and/or subjective feedbacks from a select few [15]. There are no strict guidelines on the formatting of the instruments for this objective as long as it meets the fundamental goal to improve and refine the instrument to become effective in gathering data on the study's population [11,12,23].

(b) Conduct diagnostic statistical checks on the pilot data gathered from the respondents answering the actual survey. These checks will diagnose how the respondents actually respond to each item [24] which will be further discussed in the next section.

Pilot data provides a useful, initial baseline on how the survey might perform on an actual data gathering [25]. Hence, the selected pilot respondents should possess the characteristics that are as close as possible with the actual population and are anticipated to have diverse responses on your constructs. Having a pilot respondent similar to the actual population allows an effective proxy and anticipate the potential issues that may arise when dealing with the population [15,24], while the diversity of their responses ensure that an item is capable of measuring the actual item as part of the item without suffering from response biases such as social desirability bias. While the actual population can be theoretically a gold standard for selecting pilot respondents, I recommend avoiding this approach to ensure sufficient pool of respondents for the actual data gathering. Pilot testing can be iterative, and requires a different set of pilot respondents every iteration to mitigate practice bias.

### *6.4. Statistical checks: Descriptive statistics and multicollinearity*

Assuming the pilot data has been collected, I recommend to first check and identify respondents whose responses were outlier values. This informs the pilot respondents that show unusual behavior that is deviant from the rest of the pilot respondents. This provides researchers the opportunity to probe questions on these respondents about their respondents, as well as further examine their outputs when assessing face validity. Researchers can then proceed on conducting and interpreting descriptive statistics and multicollinearity tests on the individual items, which are the primary diagnostic measures to assess how well the items are measuring the formative construct.

Descriptive statistics summarize the characteristics of a sample [27,29] and can be used in diagnosing whether the items are performing well based on the characteristics of pilot sample as the proxy estimation. Measures of central tendency such as mean and median can be used to determine the general location on their responses, informing the ability of an item to be as unbiased as possible, while measures of variability such as standard deviation and interquartile range determine the general extent of diversity or uniformity of their responses that informs the item's conceptual scope to gather various responses [26,29]. Both measures should be used to collectively inform the quality of each item which is guided from empirically established guidelines.

Given the diagnostic purpose of the pilot testing, interpreting descriptive statistics on the construct as a whole is pragmatically redundant since interpretations on central tendency and variability will still depend on the nature of the individual items because their values constantly change depending on the involved items, lacking stability of results.

Multicollinearity in the context of formative constructs occurs when two or more indicators are found to be statistically too similar that an indicator can be a numerical copy of another indicator. To measure multicollinearity, the Variance Inflation Factor (VIF) is used as the primary standard check which relies on the coefficient of determination or r-squared ($R^2$) that can be obtained by performing an Ordinary Least Squares (OLS) regression [1,3].

$$VIF = \frac{1}{1-R^2} \qquad (5)$$

However, because VIF is fundamentally computed from a regression model, it could produce unstable estimates due to low sample size that is evident in our pilot respondents. If feasible, I recommend to refer to the empirical guidelines of Hair et al. [2] on the number of observations needed to run a regression model. Even when the alternative way of assessing bivariate correlations across all possible pairs of indicators using Pearson's product-moment correlation [2], it assumes a continuous nature of the indicator data which is not always the case for Likert scales, resulting to a more serious consequence. Misspecification holds a greater risk on misleading results rather than the instability due to sample size.

The researchers can then proceed to actual data gathering only if the constructs sufficiently satisfied the conditions on the said statistical checks, as well as the preceding auxiliary procedures.

**7. CONCLUSION**

Advancing rigorous methodological practices for formative measurement models is crucial for promoting robust and valid findings since the current validation processes remains compromised and misconceived due to historical origins. Hence, this paper justifies the theoretical motivation on the perceived gaps on questionnaire validation practices and guide the pragmatic roles of the researchers in ensuring appropriate pilot testing procedures related to survey questionnaires. This addresses the long-standing misconceptions of model misspecification on formative constructs as we believe researchers should prioritize theoretical justification and contextual accuracy whenever feasible.